\documentclass[runningheads]{llncs}
\usepackage[T1]{fontenc}

\usepackage{graphicx,amsmath,amssymb,bm,subfigure,booktabs,multirow,algorithm,algorithmic,array}
\usepackage[misc]{ifsym} 
\begin{document}
\title{DULDA: Dual-domain Unsupervised Learned Descent Algorithm for PET image reconstruction}


\authorrunning{F. Author et al.}
\author{Rui Hu\inst{1,3}\and 
Yunmen Chen\inst{2}\and
Kyungsang Kim \inst{3} \and
Marcio Aloisio Bezerra Cavalcanti Rockenbach \inst{3,4} \and
Quanzheng Li\inst{3}$^{(\textrm{\Letter})}$ \and
Huafeng Liu\inst{1}$^{(\textrm{\Letter})}$
} 
\institute{State Key Laboratory of Modern Optical Instrumentation, Department of Optical Engineering,Zhejiang University, Hangzhou 310027, China
\\
\email{liuhf@zju.edu.cn} \and Department of Mathematics, University of Florida, Gainesville, FL 32611 USA
\and The Center for Advanced Medical Computing and Analysis, Massachusetts General Hospital/Harvard Medical School, Boston, MA 02114, USA\\
\email{li.quanzheng@mgh.harvard.edu}
\and Center for Clinical Data Science, Massachusetts General Brigham, Boston, MA 02116, USA
}

\titlerunning{DULDA}
\maketitle              
\begin{abstract}
Deep learning based PET image reconstruction methods have achieved promising results recently. However, most of these methods follow a supervised learning paradigm, which rely heavily on the availability of high-quality training labels. In particular, the long scanning time required and high radiation exposure associated with PET scans make obtaining these labels impractical. In this paper, we propose a dual-domain unsupervised PET image reconstruction method based on learned descent algorithm, which reconstructs high-quality PET images from sinograms without the need for image labels. Specifically, we unroll the proximal gradient method with a learnable $l_{2,1}$ norm for PET image reconstruction problem. The training is unsupervised, using measurement domain loss based on deep image prior as well as image domain loss based on rotation equivariance property. The experimental results demonstrate the superior performance of proposed method compared with maximum-likelihood expectation–maximization (MLEM), total-variation regularized EM (EM-TV) and deep image prior based method (DIP).

\end{abstract}

\section{INTRODUCTION}
Positron Emission Tomography (PET) is a widely used modality in functional imaging for oncology, cardiology, neurology, and medical research~\cite{nordberg2010use}. However, PET images often suffer from a high level of noise due to several physical degradation factors as well as the ill-conditioning of the PET reconstruction problem. As a result, the quality of PET images can be compromised, leading to difficulties in accurate diagnosis.

Deep learning (DL) techniques, especially supervised learning, have recently garnered considerable attention and show great promise in PET image reconstruction compared with traditional analytical methods and iterative methods. Among them, four primary approaches have emerged: DL-based post-denoising~\cite{cui2019pet,onishi2021anatomical}, end-to-end direct learning~\cite{zhu2018image,haggstrom2019deeppet,li2023deep}, deep learning regularized iterative reconstruction~\cite{gong2018iterative,kim2018penalized,li2022deep,li2022neural} and deep unrolled methods~\cite{mehranian2020model,lim2020improved,hu2022transem}. 

DL-based post denoising methods are relatively straightforward to implement but can not reduce the lengthy reconstruction time and its results are significantly affected by the pre-reconstruction algorithm. End-to-end direct learning methods utilize deep neural networks to learn the directing mapping from measurement sinogram to PET image. Without any physical constraints, these methods can be unstable and extremely data-hungry. Deep learning regularized iterative reconstruction methods utilize a deep neural network as a regularization term within the iterative reconstruction process to regularize the image estimate and guide the reconstruction process towards a more accurate and stable solution. 
Despite the incorporation of deep learning, the underlying mathematical framework and assumptions of deep learning regularized iterative methods still rely on the conventional iterative reconstruction methods. Deep unrolled methods utilize a DNN to unroll the iterative reconstruction process and to learn the mapping from sinogram to the reconstructed PET images, which potentially result in more accurate and explainable image reconstruction. 
Deep unrolled methods have demonstrated improved interpretabillity and yielded inspiring outcomes. 

\begin{figure}[t]           
\centering
\includegraphics[width=\textwidth]{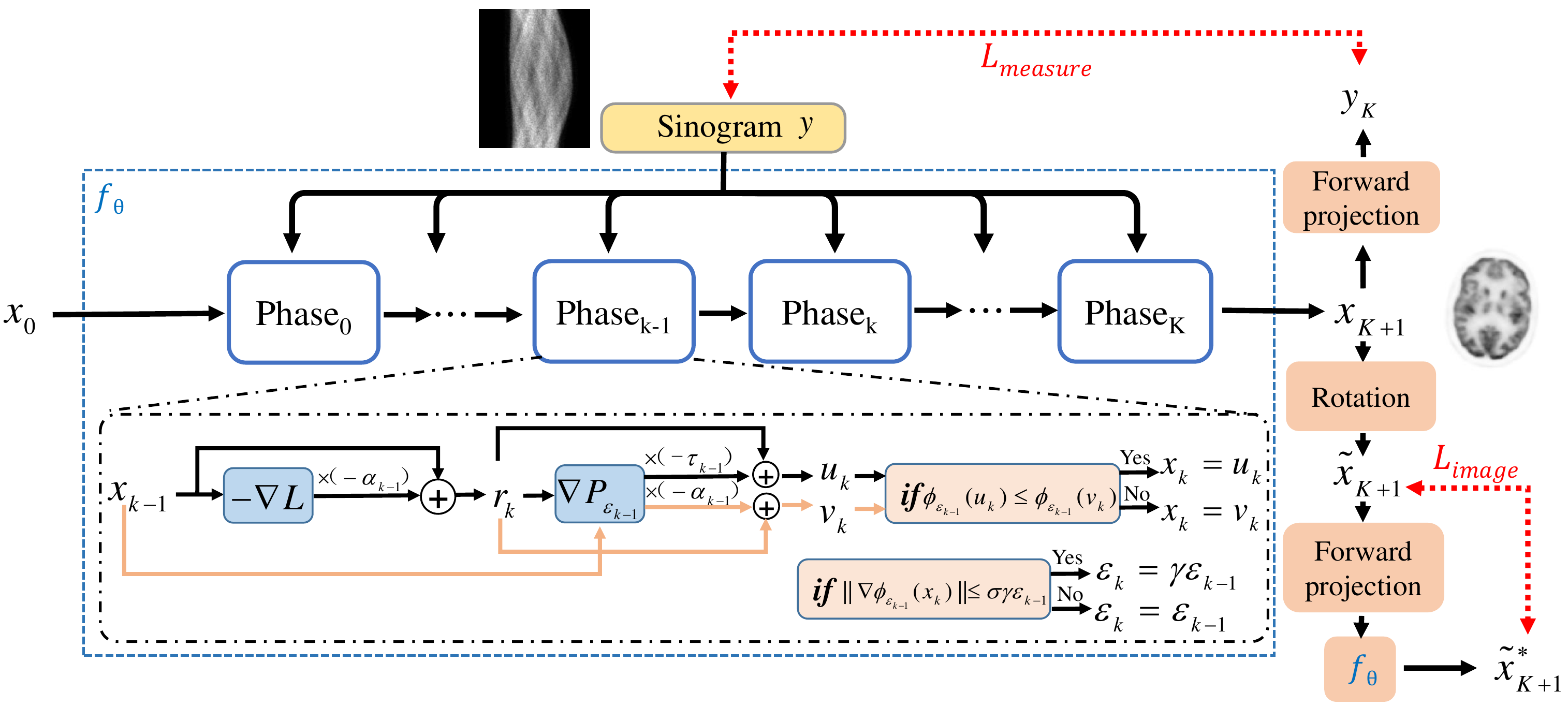}  
\caption{Diagram of the proposed DULDA for PET image reconstruction. The LDA was unrolled into several phases with the learnable $l_{2,1}$ norm, where each phase includes the gradient calculation of both likelihood and regularization.}  
\label{Fig1}                     
\end{figure}

However, the aforementioned approaches for PET image reconstruction depend on high quality ground truths as training labels, which can be difficult and expensive to obtain. This challenge is further compounded by the high dose exposure associated with PET imaging.Unsupervised/self supervised learning has gained considerable interest in medical imaging, owing to its ability to mitigate the need for high-quality training labels. Gong et al. proposed a PET image reconstruction approach using the deep image prior (DIP) framework~\cite{gong2018pet}, which employed a randomly initialized Unet as a prior. In another study, Fumio et al. proposed a simplified DIP reconstruction framework with a forward projection model, which reduced the network parameters~\cite{hashimoto2022pet}. Shen et al. proposed a DeepRED framework with an approximate Bayesian framework for unsupervised PET image reconstruction~\cite{shen2022unsupervised}.  These methods all utilize generative models to generate PET images from random noise or MRI prior images and use sinogram to design loss functions. However, these generative models 1) tend to favor low frequencies, 2) lack of mathematical interpretability, and 3) do not consider the constraints of PET physical properties. In the absence of anatomic priors, the network convergence can take a considerable amount of time, resulting in prolonged reconstruction times. Recently, equivariant property~\cite{chen2022robust} of medical imaging system is proposed to train the network without labels, which shows the potential for the designing of PET reconstruction algorithms.

In this paper, we propose a dual-domain unsupervised learned descent algorithm for PET image reconstruction, which is the first attempt to combine unsupervised learning and deep unrolled method for PET image reconstruction. The main contributions of this work are summarized as follows: 1) a novel model based deep learning method for PET image reconstruction is proposed with a learnable $l_{2,1}$ norm for more general and robust feature sparsity extraction of PET images; 2) a dual domain unsupervised training strategy is proposed, which is plug-and-play and does not need paired training samples; 3) without any anatomic priors, the proposed method shows superior performance both quantitatively and visually.

\section{METHODS AND MATERIALS}
\subsection{Problem Formulation}
As a typical inverse problem, PET image reconstruction can be modeled in a variational form and cast as an optimization task, as follows: 
\begin{equation}
\label{eq1}
    \min \phi (\bm{x};\bm{y},\bm{\theta} ) =  - L(\bm{y}|\bm{x}) + P(\bm{x};\bm{\theta} )
\end{equation}
\begin{equation}
\label{eq2}
    L(\bm{y}|\bm{x}) = \sum\limits_i {{y_i}\log {{\overline y }_i}}  - \sum\limits_i {\overline y _i}
\end{equation}
\begin{equation}
\label{eq3}
    \overline{\bm{y}} = \bm{A}\bm{x} + \bm{b}
\end{equation}
where $\bm{y}$ is the measured sinogram data, $\overline{\bm{y}}$ is the mean of the measured sinogram. $\bm{x}$ is the PET activity image to be reconstructed, $L(\bm{y}|\bm{x})$ is the Poisson log-likelihood of measured sinogram data. $P(\bm{x};\bm{\theta})$ is the penalty term with learnable parameter $\bm{\theta}$. $\bm{A} \in {{\mathbb R}^{I \times J}}$ is the system response matrix, with $A_{ij}$ representing the probabilities of detecting an emission from voxel $j$ at detector $i$.

We expect that the parameter $\bm{\theta}$ in penalty term $P$ can be learned from the training data like many other deep unrolling methods. However, most of these methods directly replace the penalty term~\cite{gong2019mapem} or its gradient~\cite{mehranian2020model,hu2022transem} with a network, which loses some mathematical rigor and interpretablities. 

\subsection{Parametric form of learnable regularization}

We choose to parameterize $P$ as the ${l_{2,1}}$ norm with a feature extraction operator $g(x)$ to be learned in the training data. The smooth nonlinear mapping $g$ is used to extract sparse features and the ${l_{2,1}}$ norm is used as a robust and effective sparse feature regularization. Specifically, we formulate $P$ as follows~\cite{chen2021learnable}:
\begin{equation}
\label{eq4}
    P(\bm{x};\bm{\theta}) = ||\bm{g}_{\bm{\theta}}(\bm{x})|{|_{2,1}} = \sum\limits_{i = 1}^m {||{\bm{g}_{i,\bm{\theta}}}(\bm{x})||}
\end{equation}
where $\bm{g}_{i,\bm{\theta}}(\bm{x})$ is $i$-th feature vector. We choose $\bm{g}$ as a multi-layered CNN with nonlinear activation function $\sigma$, and $\sigma$ is a smoothed ReLU:
\begin{equation}
\label{eq5}
\sigma (x) = \left\{ \begin{gathered}
  0,{\quad \quad\quad\quad\quad\quad\quad\text{if }}x \leqslant  - \delta , \hfill \\
  \frac{1}{{4\delta }}{x^2} + \frac{1}{2}x + \frac{\sigma }{4},{\ \text{  if}} - \delta  < x < \delta , \hfill \\
  x,{\quad\quad\quad\quad\quad\quad\quad\text{if }}x \geqslant \delta , \hfill \\ 
\end{gathered}  \right.{\text{   }}
\end{equation}
In this case, the gradient $\nabla g$ can be computed directly. The Nesterov's smoothing technique is used in $P$ for the derivative calculation of the $l_{2,1}$ norm through smooth approximation:
\begin{equation}
\label{eq6}
    {P_\varepsilon }(\bm{x}) = \sum {\frac{1}{{2\varepsilon }}} ||{\bm{g}_i}(\bm{x})|{|^2} + \sum {(||{\bm{g}_i}(\bm{x}) - \frac{\varepsilon }{2}||)}
\end{equation}
\begin{equation}
\label{eq7}
    \nabla {P_\varepsilon }(\bm{x}) = {\sum {\nabla {\bm{g}_i}(\bm{x})} ^T}\frac{{{\bm{g}_i}(\bm{x})}}{\varepsilon } + {\sum {\nabla {\bm{g}_i}(\bm{x})} ^T}\frac{{{\bm{g}_i}(\bm{x})}}{{||{\bm{g}_i}(\bm{x})||}}
\end{equation}
where parameter $\varepsilon$ controls how close the approximation $P_\varepsilon$ to the original $P$.

\begin{figure}[t]            
\centering
\includegraphics[width=\textwidth]{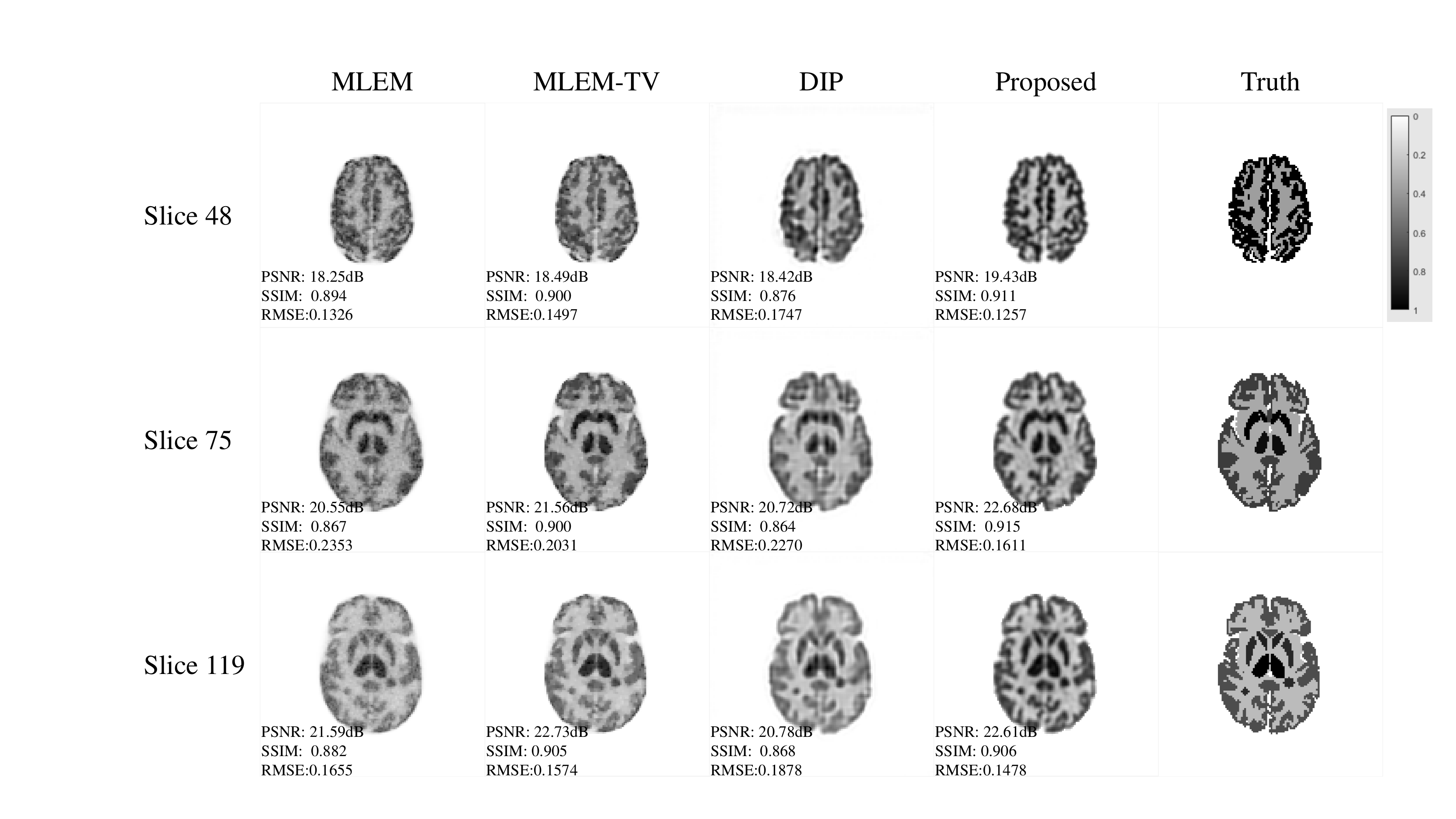}  
\caption{Reconstruction results of MLEM, EMTV, DIP and proposed DULDA on different slices of the test set.} 
\label{Fig2}
\end{figure}

\begin{algorithm}[t]
\caption{Learned Descent Algorithm for PET image reconstruction} 
\label{alg1} 
\renewcommand{\algorithmicrequire}{\textbf{Input:}}
    \begin{algorithmic}[1] 
    \REQUIRE ~Image initialization ${\bm{x_0}}$, $\rho ,\gamma  \in (0,1)$, ${\varepsilon _0},\sigma ,\tau  > 0$, maximum number of iteration $I$, total phase numbers $K$ and measured Sinogram $\bf{y}$
    \FOR{$i \in [1,I]$}
    \STATE $    {\bm{r}_k} = {\bm{x}_{k - 1}} + {\alpha _{k - 1}}(\sum {\frac{{{\bm{A}^T}\bm{y}}}{{\bm{A}{\bm{x}_{k - 1}} + \bm{b}}}}  - {\sum \bm{A} ^T \bm{1}}) $
    \STATE $    {\bm{u}_k}  = {\bm{r}_k} - {\tau _{k - 1}}{\nabla _{{P_{{\varepsilon _{k - 1}}}}}}({\bm{r}_k})$
    \REPEAT 
    \STATE $    {\bm{v}_k} = {\bm{x}_{k - 1}} - {\alpha _{k - 1}}\nabla ( - L(\bm{y}|{\bm{x}_{k - 1}})) - {\alpha _{k - 1}}\nabla {P_{{\varepsilon _{k - 1}}}}({\bm{x}_{k - 1}})$
    \UNTIL ${\phi _{{\varepsilon _{k - 1}}}}({\bm{v}_k}) \leqslant {\phi _{{\varepsilon _{k - 1}}}}({\bm{x}_{k - 1}})$
    \STATE  If ${\phi }({\bm{u}_k}) \leqslant {\phi }({\bm{v}_{k}})$, $\bm{x}_k=\bm{u}_k$; otherwise, $\bm{x}_k = \bm{v}_k$
    \STATE If $||\nabla {\phi _{{\varepsilon _{k - 1}}}}({\bm{x}_k})|| < \sigma \gamma {\varepsilon _{k - 1}}$, $\varepsilon_k = \gamma \varepsilon_{k-1}$; otherwise, $\varepsilon_k = \varepsilon_{k-1}$
    \ENDFOR
    \RETURN $\bf{x}_K$; 
    \end{algorithmic}
\end{algorithm}
\subsection{Learned Descent Algorithm for PET}
With the parametric form of learnable regularization given above, we rewrite Eq.\ref{eq1} as the objective function:
\begin{equation}
\label{eq8}
    \min \phi (\bm{x};\bm{y},\bm{\theta} ) =  - L(\bm{y}|\bm{x}) + P_\varepsilon(\bm{x};\bm{\theta} )
\end{equation}
We unrolled the learned descent algorithm in several phases as shown in Fig.\ref{Fig1}. In each phase $k-1$, we apply the proximal gradient step in Eq.\ref{eq8}:
\begin{equation}
\label{eq9}
    {\bm{r}_k} = {\bm{x}_{k - 1}} - {\alpha _{k - 1}}\nabla ( - L(\bm{y}|\bm{x})) = {\bm{x}_{k - 1}} + {\alpha _{k - 1}}(\sum {\frac{{{\bm{A}^T}\bm{y}}}{{\bm{A}{\bm{x}_{k - 1}} + \bm{b}}}}  - {\sum \bm{A} ^T \bm{1}})
\end{equation}
\begin{equation}
\label{eq10}    
    {\bm{x}_k} = {\text{pro}}{{\text{x}}_{{\alpha _{k - 1}}{P_{{\varepsilon _{k - 1}}}}}}({\bm{r}_k})
\end{equation}
where the proximal operator is defined as:
\begin{equation}
\label{eq11}
    {\text{pro}}{{\text{x}}_{\alpha P}}(\bm{r}) = \mathop {\arg \min }\limits_x \{ \frac{1}{{2\alpha }}||\bm{x} - \bm{r}|{|^2} + P(\bm{x})\}
\end{equation}
In order to have a close form solution of the proximal operator, we perform a Taylor approximation of $P_{\varepsilon_{k-1}}$:
\begin{equation}
\label{eq12}
    {\tilde P_{{\varepsilon _{k - 1}}}}(\bm{x}) = {P_{{\varepsilon _{k - 1}}}}({\bm{r}_k}) + (\bm{x} - {\bm{r}_k}) \cdot {\nabla _{{P_{{\varepsilon _{k - 1}}}}}}({\bm{r}_k}) + \frac{1}{{2{\beta _{k - 1}}}}||\bm{x} - {\bm{r}_k}|{|^2}    
\end{equation}
After discarding higher-order constant terms, we can simplify the Eq.\ref{eq10} as:
\begin{equation}
\label{eq13}
    {\bm{u}_k} = {\text{pro}}{{\text{x}}_{{\alpha _{k - 1}}{{\tilde P}_{{\varepsilon _{k - 1}}}}}}({\bm{r}_k}) = {\bm{r}_k} - {\tau _{k - 1}}{\nabla _{{P_{{\varepsilon _{k - 1}}}}}}({\bm{r}_k})
\end{equation}
where $\alpha_{k-1}$ and $\beta_{k-1}$ are two parameters greater than 0 and ${\tau _{k - 1}} = \frac{{{\alpha _{k - 1}}{\beta _{k - 1}}}}{{{\alpha _{k - 1}} + {\beta _{k - 1}}}}$. We also calculate a close-form safeguard $\bm{v}_k$ as:
\begin{equation}
\label{eq14}
    {\bm{v}_k} = {\bm{x}_{k - 1}} - {\alpha _{k - 1}}\nabla ( - L(\bm{y}|{\bm{x}_{k - 1}})) - {\alpha _{k - 1}}\nabla {P_{{\varepsilon _{k - 1}}}}({\bm{x}_{k - 1}})
\end{equation}
The line search strategy is used by shrinking $\alpha_{k-1}$ to ensure objective function decay. We choose the $\bm{u}_k$ or $\bm{v}_k$ with smaller objection function value $\phi_{\varepsilon_{k-1}}$ to be the next $\bm{x}_k$. The smoothing parameter $\varepsilon_{k-1}$ is shrinkage by $\gamma  \in (0,1)$ if the $||\nabla {\phi _{{\varepsilon _{k - 1}}}}({\bm{x}_k})|| < \sigma \gamma {\varepsilon _{k - 1}}$ is satisfied.
The whole flow is shown in Algorithm \ref{alg1}.

\begin{figure}[t]            
\centering
\includegraphics[width=\textwidth]{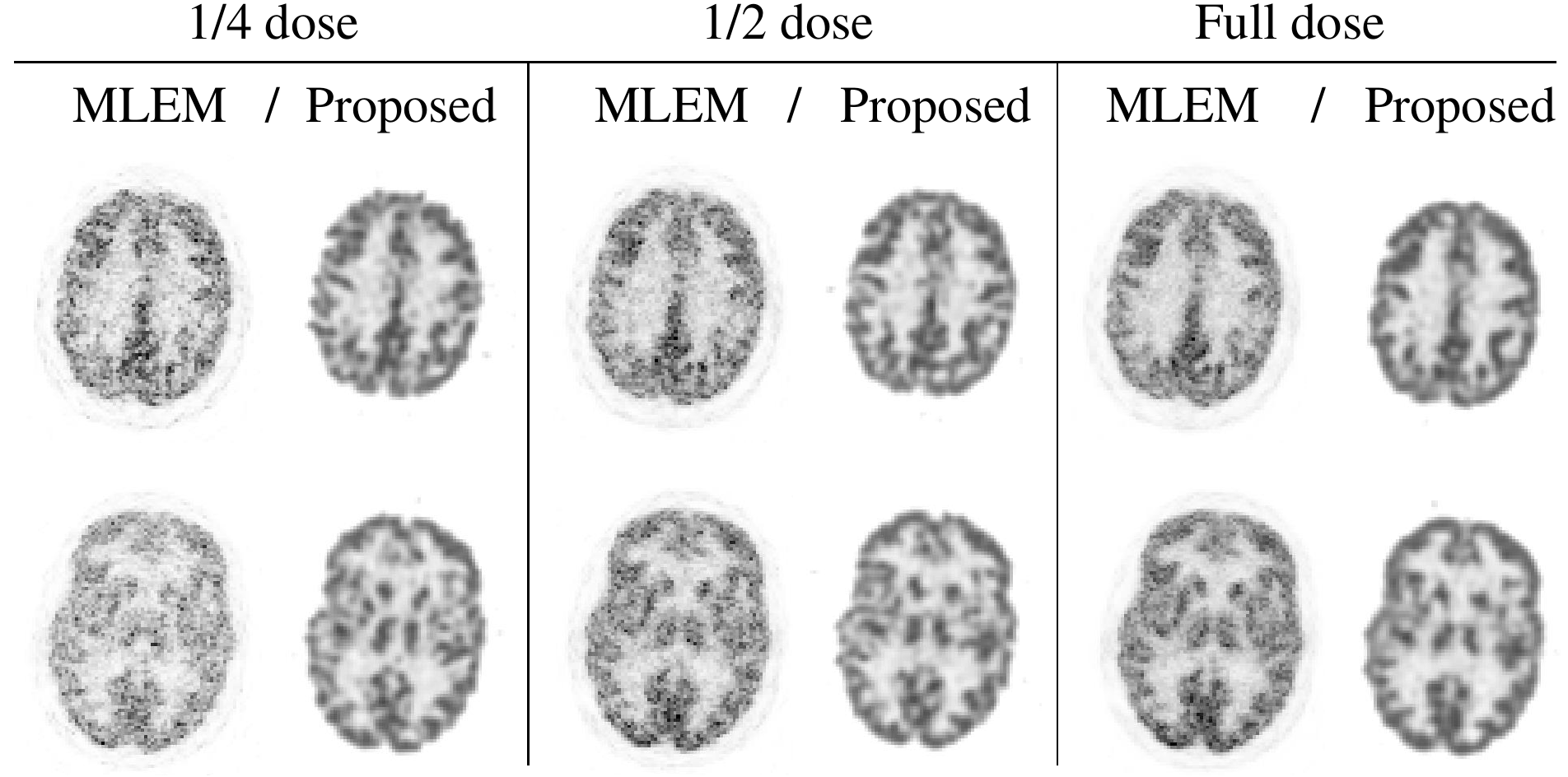}  
\caption{The robust analysis on proposed DULDA with one clinical patient brain sample with different dose level. From left to right: MLEM results and DULDA results with quarter dose sinogram, half dose sinogram and full dose sinogram.} 
\label{Fig3}
\end{figure}

\subsection{Dual-Domain Unsupervised Training}
The whole reconstruction network is indicated by $f_\theta$ with learned parameter $\theta$. Inspired by Deep image prior~\cite{ulyanov2018deep} and equivariance~\cite{chen2022robust} of PET imaging system, the proposed dual-domain unsupervised training loss function is formulated as:
\begin{equation}
\label{eq15}
    {L_{dual}} = {L_{image}} + \lambda {L_{measure}}
\end{equation}
where $\lambda$ is the parameter that controls the ratio of different domain loss function, which was set to 0.1 in the experiments. For image domain loss $L_{image}$, the equivariance constraint is used. For example, if the test sample $x_t$ first undergoes an equivariant transformation, such as rotation, we obtain $x_{tr}$. Subsequently, we perform a PET scan to obtain the sinogram data of $x_{tr}$ and $x_t$. The image reconstructed by the $f_\theta$ of these two sinogram should also keep this rotation properties. The $L_{image}$ is formulate as:
\begin{equation}
\label{eq16}
{L_{image}} = ||{T_r}\underbrace {{f_\theta }(\bm{y})}_{{x_t}} - {f_\theta }(A(\underbrace {{T_r}({f_\theta }(\bm{y}))}_{{x_{tr}}}))|{|^2}
\end{equation}
where $T_r$ denotes the rotation operator, $A$ is the forward projection which also can be seen as a measurement operator.
For sinogram domain loss $L_{measure}$, the data argumentation with random noise $\bm{\xi}$ is performed on $\bm{y}$:
\begin{equation}
\label{eq17}
{L_{measure}} = ||(\bm{y} + \bm{\xi} ) - A{f_\theta }(\bm{y} + \bm{\xi} )|{|^2}
\end{equation}

\subsection{Implementation details and reference methods}

We implemented DULDA using Pytorch 1.7 on a NVIDIA GeForce GTX Titan X. The Adam optimizer with a learning rate of $10^{-4}$ was used and trained for 100 epochs with batch size of 8. The total unrolled phase was 4. The image $x_0$ was initialized with the values of one. The smoothing parameter $\varepsilon_0$ and $\delta$ were initialized to be 0.001 and 0.002. The step-size $\alpha_0$ and $\beta_0$ were initialized to be 0.01 and 0.02. The system matrix was computed by using Michigan Image Reconstruction Toolbox (MIRT) with a strip-integral model~\cite{noh2009statistical}. The proposed DULDA was compared with MLEM~\cite{shepp1982}, total variation regularized EM (EM-TV)~\cite{jonsson1998total} and deep image prior method (DIP)~\cite{hashimoto2022pet}. For both MLEM and EM-TV, 25 iterations were adopted. The penalty parameter for EM-TV was $2e^{-5}$. For DIP, we used random noise as input and trained 14000 epochs with the same training settings as DULDA to get the best results before over-fitting.

\begin{table}[t]
\centering
\caption{Quantitative analysis and bias-variance analysis for the reconstruction results of MLEM, EM-TV, DIP and Proposed DULDA.}
\resizebox{\textwidth}{!}{
\begin{tabular}{@{}ccccccc@{}}
\toprule
Methods  \qquad& PSNR(dB)$\uparrow$     \qquad& SSIM$\uparrow$          \qquad& RMSE$\downarrow$         \qquad& CRC$\uparrow$     \qquad& Bias$\downarrow$   \qquad& Variance$\downarrow$ \\ \midrule
MLEM     \qquad& 20.02 ±1.91 \qquad& 0.889 ±0.015 \qquad& 0.160 ±0.045 \qquad& 0.6517 \qquad& 0.5350 \qquad& 0.2311   \\
EM-TV    \qquad& 20.28 ±2.21 \qquad& 0.904 ±0.014 \qquad& 0.154 ±0.044 \qquad& 0.8027 \qquad& 0.5389 \qquad& 0.2340   \\
DIP      \qquad& 19.96 ±1.50 \qquad& 0.873 ±0.012 \qquad& 0.187 ±0.047 \qquad& 0.8402 \qquad& 0.2540 \qquad& \pmb{0.2047}   \\
Proposed \qquad& \pmb{20.80 ±1.77} \qquad& \pmb{0.910 ±0.011} \qquad& \pmb{0.148 ±0.011} \qquad& \pmb{0.8768} \qquad& \pmb{0.2278} \qquad& 0.2449   \\ \bottomrule
\end{tabular}
}
\label{table1}
\end{table}

\begin{table}[t]
\caption{Ablation study for different phase numbers and loss function type of DULDA on the test datasets.}
\centering
\setlength{\tabcolsep}{3mm}{
\begin{tabular}{@{}|cc|c|c|c|@{}}
\hline
\multicolumn{2}{|c|}{Settings}                              &  PSNR$\uparrow$        & SSIM$\uparrow$         & MSE$\downarrow$          \\ \hline
\multicolumn{1}{|c|}{\multirow{5}{*}{phase   numbers}} & 2  & 14.53 ±1.45 & 0.769 ±0.024 & 0.314 ±0.047 \\ 
\multicolumn{1}{|c|}{}                                 & 4  & 20.80 ±1.77 & 0.910 ±0.011 & 0.148 ±0.011 \\ 
\multicolumn{1}{|c|}{}                                 & 6  & 20.29 ±1.16 & 0.903 ±0.014 & 0.156 ±0.016 \\ 
\multicolumn{1}{|c|}{}                                 & 8  & 19.94 ±1.31 & 0.884 ±0.012 & 0.180 ±0.013 \\ 
\multicolumn{1}{|c|}{}                                 & 10 & 15.33 ±0.65 & 0.730 ±0.020 & 0.313 ±0.050 \\ \hline
\multicolumn{2}{|c|}{only $L_{image}$}                         & 15.41 ±0.69 & 0.729 ±0.008 & 0.324 ±0.048 \\ \hline
\multicolumn{2}{|c|}{only $L_{measure}$}                       & 19.61 ±1.49 & 0.881 ±0.012 & 0.181 ±0.011 \\ \hline
\multicolumn{2}{|c|}{$L_{image}$ $+$ $L_{measure}$}                 & 20.80 ±1.77 & 0.910 ±0.011 & 0.148 ±0.011 \\ \hline
\end{tabular}}

\label{tab2}
\end{table}
\section{EXPERIMENT AND RESULTS}
\subsection{Experimental evaluations}
Forty 128$\times$128$\times$40 3D Zubal brain phantoms~\cite{zubal1994computerized} were used in the simulation study as ground truth, and one clinical patient brain images with different dose level were used for the robust analysis. Two tumors with different size were added in each Zubal brain phantom. The ground truth images were firstly forward-projected to generate the noise-free sinogram with count of $10^6$ for each transverse slice and then Poisson noise were introduced. 20 percent of uniform random events were simulated. In total, 1600 (40$\times$40) 2D sinograms were generated. Among them, 1320 (33 samples) were used in training, 200 (5 samples) for testing, and 80 (2 samples) for validation. A total of 5 realizations were simulated and each was trained/tested independently for bias and variance calculation~\cite{gong2018pet}. We used peak signal to noise ratio (PSNR), structural similarity index (SSIM) and root mean square error (RMSE) for overall quantitative analysis. The contrast recovery coefficient (CRC)~\cite{qi1999theoretical} was used for the comparison of reconstruction results in the tumor region of interest (ROI) area. 
\subsection{Results}
Fig. \ref{Fig2} shows three different slices of the reconstructed brain PET images using different methods. The DIP method and proposed DULDA have lower noise compared with MLEM and EM-TV visually. However, the DIP method shows unstable results cross different slices and fails in the recovery of the small cortex region. The proposed DULDA can recover more structural details and the white matter appears to be more sharpen. The quantitative and bias-variance results are shown in Table \ref{table1}. We noticed that DIP method performs even worse than MLEM without anatomic priors. The DIP method demonstrates a certain ability to reduce noise by smoothing the image, but this leads to losses in important structural information, which explains the lower PSNR and SSIM. Both DIP method and DULDA have a better CRC and Bias performance compared with MLEM and EM-TV.

\section{DISCUSSION}
To test the robustness of proposed DULDA, we forward-project one patient brain image data with different dose level and reconstructed it with the trained DULDA model. The results compared with MLEM are shown in Fig.\ref{Fig3}. The patient is scanned with a GE Discovery MI 5-ring PET/CT system. The real image has very different cortex structure and some deflection compared with the training data. It can be observed that DULDA achieves excellent reconstruction results in both details and edges across different dose level and different slices.Table \ref{tab2} shows the ablation study on phase numbers and loss function for DULDA. It can be observed that the dual domain loss helps improve the performance and when the phase number is 4, DULDA achieves the best performance.

\section{CONCLUSIONS}
In this work, we proposed a dual-domain unsupervised model-based deep learning method (DULDA) for PET image reconstruction by unrolling the learned descent algorithm. Both quantitative and visual results show the superior performance of DULDA when compared to MLEM, EM-TV and DIP based method. Future work will focus more on clinical aspects.

\end{document}